\documentclass[aps,prl,superscriptaddress,twocolumn,floatfix,showpacs]{revtex4-2}
\usepackage[utf8]{inputenc}
\usepackage[T1]{fontenc}
\usepackage{graphicx,color}
\usepackage{makecell}
\usepackage[utf8]{inputenc}
\usepackage{amsmath,,amssymb,bm,amsfonts,dsfont,mathrsfs,amsthm}
\usepackage{braket}
\usepackage{txfonts,comment}
\usepackage[colorlinks=true,linkcolor=blue,citecolor=blue,urlcolor=blue]{hyperref}
\usepackage{soul}
\usepackage{tikz}
\usetikzlibrary{quantikz2}
\usetikzlibrary{shapes, arrows, positioning, calc, backgrounds}
\addtocontents{toc}{\protect\setcounter{tocdepth}{-1}}

\hypersetup{colorlinks=true, linkcolor=blue, citecolor=blue, urlcolor=blue}

\begin{document}

\addtocontents{toc}{\protect\setcounter{tocdepth}{-1}}

\title{Quantum error correction for multiparameter metrology}

\author{Mauricio Guti\'errez}
\affiliation{Escuela de Química, Universidad de Costa Rica, San José 2060, Costa Rica}
\author{Chiranjib Mukhopadhyay}
\affiliation{Institute of Fundamental and Frontier Sciences, University of Electronic Sciences and Technology of China, Chengdu 611731, China}
\affiliation{Key Laboratory of Quantum Physics and Photonic Quantum Information, Ministry of Education, University of Electronic Science and Technology of China, Chengdu 611731, China}
\thanks{CM and VM contributed equally}
\author{Victor Montenegro}
\affiliation{College of Computing and Mathematical Sciences, Department of Applied Mathematics and Sciences, Khalifa University of Science and Technology, 127788 Abu Dhabi, United Arab Emirates}
\affiliation{Institute of Fundamental and Frontier Sciences, University of Electronic Sciences and Technology of China, Chengdu 611731, China}
\affiliation{Key Laboratory of Quantum Physics and Photonic Quantum Information, Ministry of Education, University of Electronic Science and Technology of China, Chengdu 611731, China}
\thanks{CM and VM contributed equally}

\author{Abolfazl Bayat}
\affiliation{Institute of Fundamental and Frontier Sciences, University of Electronic Sciences and Technology of China, Chengdu 611731, China}
\affiliation{Key Laboratory of Quantum Physics and Photonic Quantum Information, Ministry of Education, University of Electronic Science and Technology of China, Chengdu 611731, China}
\affiliation{Shimmer Center, Tianfu Jiangxi Laboratory, Chengdu 641419, China}



\begin{abstract}
For single-parameter sensing, Greenberger-Horne-Zeilinger (GHZ) probes achieve optimal quantum-enhanced precision across the unknown parameter range,  solely relying on parameter-independent separable measurement strategies for all values of the unknown parameter. 
However, in the multiparameter setting,  a single GHZ probe not only fails to achieve quantum advantage but also the corresponding optimal measurement becomes complex and dependent on the unknown parameters. Here, we provide a recipe for multiparameter sensing with GHZ probes using quantum error correction techniques by treating all but one  unknown parameters as noise, whose effects can be corrected.   This strategy restores the core advantage of single parameter GHZ-based quantum sensing, namely reaching optimally quantum-enhanced precision for all unknown parameter values while keeping the measurements separable and fixed. Specifically, given one shielded  ancilla qubit per GHZ probe, our protocol extracts optimal possible precision for any probe size. While this optimal precision is shot-noise limited for a single GHZ probe, we recover the Heisenberg scaling through use of multiple complementary GHZ probes. We demonstrate the effectiveness of the protocol with Bayesian estimation. 
\end{abstract}
\date{\today}
\maketitle

\emph{Introduction.--} Exploiting quantum features allows sensing precision beyond the capability of classical probes. This quantum enhancement of sensitivity underlies the development of quantum sensing as a key contemporary quantum technology~\cite{helstrom1969quantum,paris2004quantum, giovannetti2006quantum,paris2009quantum,schnabel2010quantum, giovannetti2011advances,toth2014quantum, degen2017quantum,Liu2020,meyer2021fisher, montenegro2025review}. Quantum sensors can be divided into two broad categories: (i) interferometric; and (ii) many-body sensors. For interferometric-based sensing, the parameter to estimate is encoded through a phase shift acting on the quantum state of a probe~\cite{Dowling_2008,Kacprowicz_2010,aasi2013enhanced, cooper2012robust,lang2013optimal,lang2014optimal,humphreys2013quantum, ou2012enhancement,Genovese_2021}. The paradigmatic example is a probe initialized in a Greenberger-Horne-Zeilinger (GHZ) state, whose collective phase shift can be efficiently measured with a precision that scales quadratically with probe size \cite{giovannetti2006quantum,belliardo2020achieving,che2019multi,Hayashi_2022,chen2025achieving}. This genuine quantum enhancement is termed Heisenberg scaling~\cite{Kok_2004,roy2008exponentially,boixo2008generalized,pezze2018quantum,Pirandola_2018} to distinguish it from classically attainable linear scaling of precision known as shot-noise  scaling (also called quantum standard limit)~\cite{Blanter_2000}. 
For many-body sensors, quantum-enhanced sensitivity is achievable around various criticalities~\cite{zanardi2008quantum,invernizzi2008quantum,rams2018at,agarwal2025quantum,ye2024essay,garbe2020critical,sarkar2022free,montenegro2021global,sahoo2024localization, alushi2024optimality,Alushi_2025,mukhopadhyay2024modular,montenegro202sequential,salvia2023critical, criticality2025beaulieu,montenegro2021global,mukhopadhyay2025saturable,abiuso2025fundamental,Mukhopadhyay_2025,puig2025from,mihailescu2025critical,Chattopadhyay2025sensing} (see Ref.~\cite{montenegro2025review} for a comprehensive review). However, such enhancement is restricted to a narrow region around the phase transition point and generally requires complex entangled measurements. In contrast, GHZ-based sensors possess two significant advantages for single-parameter sensing: (i) quantum-enhanced precision is obtained for any parameter range; and (ii) optimal measurement is always separable between qubits, and independent of the unknown parameter. Crucially, however, both of these advantages disappear for multiparameter sensing due to incompatibility of optimal measurements required to estimate different parameters \cite{ragy2016compatibility, Liu2020,Demkowicz_Dobrza_ski_2020,Razavian_2020,Candeloro_2021,pezz2025advances,mihailescu2025uncertain,mukhopadhyay2025beating,mihailescu2025metrological}. Thus a key open question is -- can we design a sensing protocol to restore these advantages to multiparameter sensing within the GHZ paradigm?



On the other hand, quantum error correction (QEC)~\cite{ShorCode,TheoryQECC, Shor96_1,Calderbank97,KitaevAnyons,TerhalReviewQEC} has emerged as a crucial tool for achieving large-scale fault-tolerant quantum computation, as it provides the clearest route for successfully running deep algorithms with a quantum advantage~\cite{ShorFactoring,Lloyd96,Alan2005,grover1996fast,kitaev1995quantum,harrow2009quantum}.  A wide variety of QEC schemes have been proposed, ranging from repetition codes~\cite{eczoo_quantum_repetition} to concatenated codes~\cite{ConcatCodes}, topological codes~\cite{fowler2012surface,bombin2006color}, and bosonic codes~\cite{gottesman2001encoding}, to name a few.  In all of these protocols, several physical qubits are used to construct a logical qubit which allows the detection and subsequent correction of certain errors arising from control imperfections and environmental interactions. This capability has sparked the adaptation of QEC to several other domains of quantum technologies, such as quantum communication~\cite{ekert1996quantum,braunstein1998quantum} and quantum sensing~\cite{dur2014improved, kessler2014quantum,unden2016quantum,martinis2015qubit,ouyang2021robust, zhou2018achieving,layden2019ancilla, Gorecki2020optimalprobeserror,zhou2020optimal,chen2024quantum,mann2025quantum,kwon2025restoring}. In the quantum sensing context, QEC has been proposed for removing unwanted environmental noise which degrades sensing precision of a target parameter. Beyond this established role of correcting environmental noise to preserve quantum enhancement, the role of QEC for multiparameter metrology is less explored. This naturally leads to the question -- can we harness QEC for GHZ-based multiparameter metrology so that one restores their key advantage of Heisenberg scaling reachable through fixed and separable measurements for all parameter values?  \\

In this letter, we answer this question in the affirmative by exploiting QEC to reduce the multiparameter estimation problem into a single-parameter one -- treating other unknown parameters as noise acting on the probe and correcting for their effects. First, we formulate a sensing protocol incorporating QEC for GHZ probes with a fixed and separable measurement strategy. Second, we show that by introducing a single shielded ancilla qubit, our protocol extracts the optimally achievable, albeit shot-noise limited, precision from a arbitrarily large single GHZ probe. Third, we optimally reach desired Heisenberg scaling using this ancilla-assisted protocol with complementary GHZ probes. We further demonstrate the effectiveness of this protocol with a Bayesian simulation. \\

\begin{figure*}[t]
    \centering
    \includegraphics[width=0.8\linewidth]{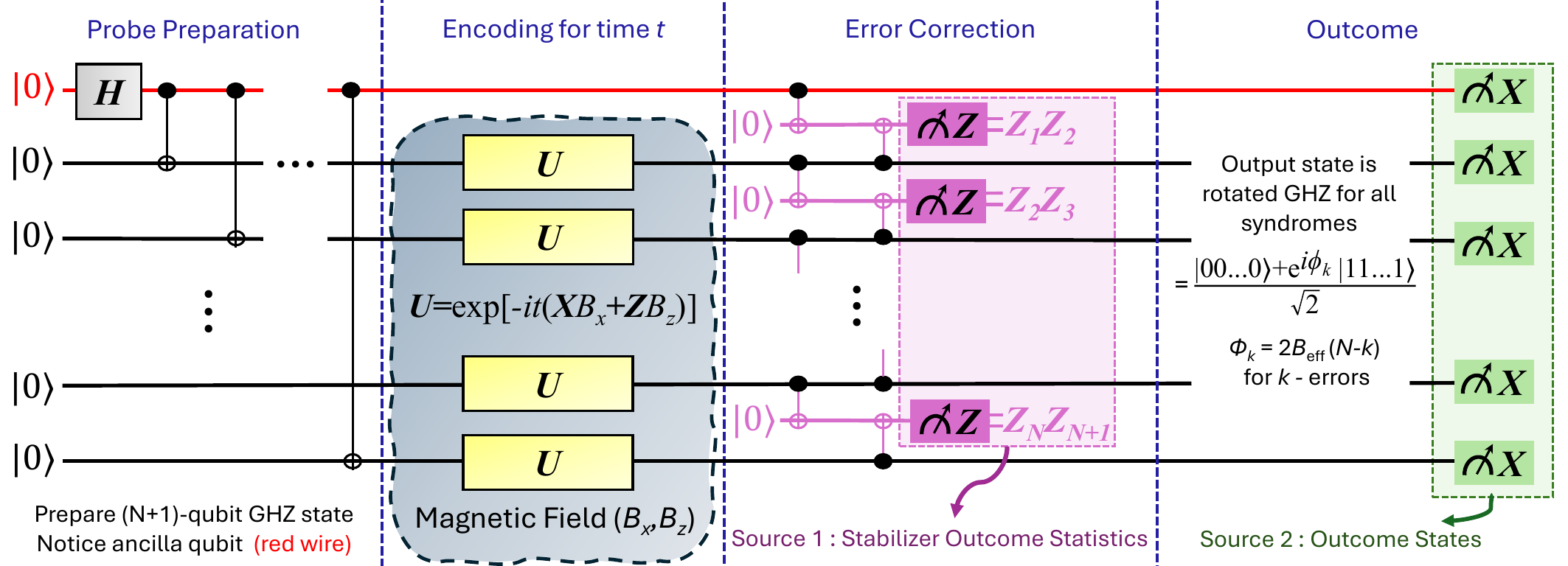}
    \caption{\textbf{Schematic of the protocol (ancilla-assisted single-probe $|\textrm{GHZ}_Z \rangle$ variant):}  The probe is prepared in the state $|\textrm{GHZ}_Z \rangle \propto |0 \rangle ^{\otimes (N+1)} + |1 \rangle ^{\otimes (N+1)}$ via an initial Hadamard gate and chained CNOTs. After all but one of the probe's qubits interact with the magnetic field, QEC is performed by measuring the stabilizers of the $(N+1)$-qubit bit-flip code.  After applying the appropriate correction and returning the state to the codespace, the $X$-component of the magnetic field is completely suppressed and the final state is proportional to $|0 \rangle ^{\otimes (N+1)} + e^{i\phi_k} |1 \rangle ^{\otimes (N+1)}$.  The acquired relative phase angle will depend on the number of detected $X$ errors, $k$, as well as $B_x$ and $B_z$ (since $\phi_k = 2 B_\textrm{eff} (N-k)$ and $B_\textrm{eff} = \arctan[B_z \tan(Bt)/B]$).  Estimating $\phi_k$ corresponds to a single-parameter quantum sensing problem for which the optimal measurement basis is known to be $X$.  We can then combine this protocol with an analogous one employing the phase-flip code to achieve Heisenberg scaling in the joint estimation of the two unknown parameters, $B_x$ and $B_z$.}
    \label{fig:schematic}
\end{figure*}

\emph{Multi-parameter quantum sensing.--} We consider a probe in the quantum state $|\psi (\vec{B})\rangle$, where $\vec{B}{=}B_x \hat{n}_x{+}B_z\hat{n}_z{\equiv}(B_x{,}B_z)$ is an unknown two-dimensional magnetic field. The sensing precision is quantified via covariance matrix  $\textrm{Cov}(\mathfrak{B})$ with elements $[\textrm{Cov}(\mathfrak{B})]_{a,b}{=}\langle \mathfrak{B}_a \mathfrak{B}_b \rangle {-} \langle \mathfrak{B}_a\rangle\langle \mathfrak{B}_b\rangle$ ($a,b{=}x,z$)  where $\mathfrak{B}{=}(\mathfrak{B}_x,\mathfrak{B}_z)$ is the estimator of the magnetic field.  The covariance matrix is bounded by the Cram\'er-Rao bound (CRB) as \cite{cramer1999mathematical,rao1945information,Liu2020,montenegro2025review} 
\begin{equation} 
     \textrm{Cov}(\mathfrak{B}) \geq \frac{F^{-1}}{M} \geq \frac{Q^{-1}}{M}     
     \label{eq:crb}
\end{equation}
where $M$ is the number of measurement samples,  $F$ is the Classical Fisher Information Matrix (CFIM) and $Q$ is the Quantum Fisher Information Matrix (QFIM). For a given  measurement setting, described by a set of Positive Operator-Valued Measure (POVM) $\Pi{=}\lbrace\Pi_i\rbrace$, the bound is given by CFIM, which is defined elementwise as 
\begin{equation}
[F]_{a,b}{=}\sum_{i} \left(\frac{\partial p_i}{\partial B_a}\frac{\partial p_i}{\partial B_b}\right)/p_i \quad ; \quad p_i{=}\bra{\psi(\vec{B})}\Pi_i\ket{\psi(\vec{B})}
\label{eq:cfi}
\end{equation}

Optimizing over all possible measurement settings $\Pi$ yields the second bound given by the QFIM whose elements are defined as $[Q]_{a,b}{=}4 \textrm{Re} \left[ \, \langle \partial_{B_a} \psi | \partial_{B_b} \psi \rangle - \langle \partial_{B_a} \psi | \psi \rangle \langle \psi |\partial_{B_b} \psi \rangle \, \right]$. While the CRB in Eq.~(\ref{eq:crb}) is in a matrix form, it can be recast as a scalar bound by multiplying both sides with a weight matrix $W$ and computing the trace. In particular, choosing $W{=}\mathbb{I}_2$ yields the bound $\textrm{Var}[\mathfrak{B}_x]{+}\textrm{Var}[\mathfrak{B}_z] {\geq} \frac{\textrm{Tr}[F^{-1}]}{M} {\geq} \frac{\textrm{Tr}[Q^{-1}]}{M}$, which we use for all subsequent analysis. \\

\emph{The model.--} We assume a quantum probe with $N$-qubit initialized at $|\psi_0\rangle$. The system evolves under the action of the two-dimensional magnetic field vector $\vec{B}$, whose effect is described by the Hamiltonian 
$\hat{H}{=}\sum_{i=1}^N \left(B_x \hat{X}_i{+}B_z \hat{Z}_i\right)$, where $\hat{X}$ and $\hat{Z}$ are the Pauli spin matrices. We denote the total magnitude of the magnetic field as $B{=}(B_x^2{+}B_z^2)^{1/2}$. After letting the probe evolve for time $t$, its quantum state is described as $\ket{\psi_t(\vec{B})}=\hat{U}^{\otimes N}\ket{\psi_0}$, where $\hat{U}{=}\exp[-it(B_x\hat{X}+B_z\hat{Z})]$ is the unitary evolution acting on a single qubit, given elementwise in the computational basis as 
\begin{equation}
U = \begin{pmatrix}
    u_{00} & u_{01} \\
    u_{01} & u_{11}
\end{pmatrix} = \begin{pmatrix}
    \cos Bt {-} i B_z\sin Bt/B & -iB_x\sin Bt/B \\\
    -iB_x \sin Bt/B & \cos Bt {+} iB_z\sin Bt/B
\end{pmatrix}
\label{eq:single_qubit_rotation}
\end{equation}  

\noindent Note that,
in the case of single parameter sensing where $B_x{=}0$, one can initialize the probe in  $\ket{\psi_0}{=}\ket{\textrm{GHZ}_Z}$, where  $|\textrm{GHZ}_Z\rangle{=}\left (|0 \rangle ^{\otimes N} {+} |1 \rangle ^{\otimes N} \right) {/} \sqrt{2}$, to  achieve quantum-enhanced sensitivity for all values of the unknown parameter using a fixed separable measurement basis. Symmetrically, if $B_z{=}0$, then the same can be achieved by $\ket{\psi_0}{=}\ket{\textrm{GHZ}_X}$, where  $|\textrm{GHZ}_X\rangle{=}\left (|+ \rangle ^{\otimes N} {+} |- \rangle ^{\otimes N} \right) {/} \sqrt{2}$ with $\ket{\pm}{=}(\ket{0}{\pm}\ket{1}){/}\sqrt{2}$.  In fact, if the direction of the magnetic field, i.e. $B_z/B_x$, is known then the problem is reduced to a single-parameter sensing for which the optimal initial state is the GHZ in the direction of the field. In the case of two-parameter sensing, where both $B_x$ and $B_z$ are unknown, then the optimal initial state is not uniquely defined and both the optimal measurement as well as the sensing precision depend on the value of $(B_x,B_z)$. \\

\emph{Error-corrected multi-parameter magnetometry.--} We start with a probe initialized in $\ket{\psi_0}{=}\ket{\mathrm{GHZ}_Z}$. Under the evolution of Hamiltonian $\hat{H}$ defined earlier, one encodes $\vec{B}$ into the quantum state $\ket{\psi_t(\vec{B})}$. The ultimate precision bound for estimating $\vec{B}$ is given in terms of the QFIM $Q$ of  the quantum state $\ket{\psi_t(\vec{B})}$. However, the measurement setting required to reach this precision bound is typically highly entangled and in principle requires perfect prior knowledge of parameter values $(B_x,B_z)$ themselves. In contrast, our goal with the error-corrected protocol is to devise a simple and fixed strategy that optimally retrieves sensing information for all unknown parameter values. Note that the initial probe state $|\textrm{GHZ}_Z\rangle$ can be visualized as a logical qubit state $|+\rangle_L {=} \left (|0 \rangle_L {+} |1 \rangle_L \right) {/} \sqrt{2}$ of the $N$-qubit bit-flip code \cite{RepetitionCode}, where $|0 \rangle_L {\equiv} |0 \rangle ^{\otimes N}, |1 \rangle_L {\equiv} |1 \rangle ^{\otimes N}$. The dynamics is a periodic precession of this effective logical qubit along the rotation axis given by the magnetic field direction.  Since the bit-flip code is designed to correct for bit-flips ($X$ errors), we treat the magnetic field in the $X$-direction as an unwanted interaction whose effect can be removed from the quantum state by performing QEC.  The stabilizer group~\cite{GottesmanThesis}, i.e., the group of all parity check operators, of the bit-flip code is generated by $S {=} \{ \hat{Z}_1\hat{Z}_2 \, , \, \hat{Z}_2\hat{Z}_3 \, , \, \ldots \, , \hat{Z}_{N-1}\hat{Z}_N \}$.  Once the information is encoded in the probe, one has to perform non-demolition stabilizer measurements. This can be achieved by using additional qubits, as shown in Fig.~\ref{fig:schematic}.  Measuring the stabilizers plays two key roles.  First, it transforms the $X$-component of the continuous rotation $\hat{U}^{\otimes N}$ into discrete, stochastic bit-flips.  Second, the stabilizer measurement outcomes, i.e., syndromes, indicate the error subspace where the state has been projected and determine the required correcting operation (see appendix for a concrete illustration) to return to the codespace spanned by $\left \{|0\rangle_L, |1\rangle_L \right \}$. The syndromes corresponding to a given number of bit flip errors, say $k$,  all appear with equal probabilities regardless of error location. We thus group them together to obtain a total probability $p_k$ for $k$ errors, where $k {\in} \left\{ \, 0, \, \ldots \,, \, \lfloor (N{-}1){/}2 \rfloor \, \right\}$ (see appendix for explicit form of $p_k$). One may try to estimate $(B_x,B_z)$, relying only on these measurement outcomes. However, the corresponding CFIM $F^{\rm stab}$, computed from outcome statistics $\{p_k\}$ of the stabilizer measurement outcomes using Eq.~(\ref{eq:cfi}), is singular. In other words, stabilizer measurement outcome statistics alone is not informative enough to estimate the unknown field $\vec{B}$, and we need additional information. 

The combined action of the unitary rotation given by the magnetic field, the projective stabilizer measurements, and the correction back to the codespace results in an effective logical single-qubit operation. The effective rotation of the logical qubit basis for each detected $X$ error of weight $k$ is then given by
\begin{eqnarray}
    & |0 \rangle_L \longrightarrow  |\tilde{0} \rangle_L \propto \, u_{00}^{N-k}u_{01}^k \, | 0 \rangle_L \, + \, u_{01}^{N-k}u_{11}^k \, |1 \rangle_L \label{eq:zerolog} \nonumber \\
    & |1 \rangle_L \longrightarrow  |\tilde{1} \rangle_L\, \propto \, u_{01}^{N-k}u_{11}^k \, | 0 \rangle_L \, + \, u_{01}^k u_{11}^{N-k} \, |1 \rangle_L,
\label{eq:ecvm_rotation}
\end{eqnarray}
\noindent where $u$ are the relevant matrix elements from Eq.~\eqref{eq:single_qubit_rotation}, and $k {\in} \left\{ \, 0, \, \ldots \,, \, \lfloor (N{-}1){/}2 \rfloor \, \right\}$. Consequently, the unnormalized  post-error-correction quantum state, after detecting $k$ errors is
\begin{equation}\label{eq:PEC}
    \ket{\psi_{k}^{\rm PEC}}{\sim} \left(u_{00}^{N-k}u_{01}^k {+} u_{01}^{N-k}u_{11}^k\right)\ket{0}^{\otimes N}{+} \left((u_{01}^{N-k}u_{11}^k {+} u_{01}^k u_{11}^{N-k}\right)\ket{1}^{\otimes N}
\end{equation}
\textcolor{black}{Note that although the above post-error-correction quantum state is in the subspace of the logical qubits, the coefficients of $\ket{0}_L$ and  $\ket{1}_L$, in general, are not equal. Therefore, the quantum state $\ket{\psi_{k}^{\rm PEC}}$ is not in the form of a phase-shifted GHZ state $(\ket{0}_L {+} e^{i\phi_k}|1\rangle_L)/\sqrt{2}$ (see appendix for explicit expression). Nonetheless, for benchmarking reasons which will become clear later, we still measure the string operator $\prod_k \hat{X}_k$ (not necessarily optimal), whose two measurement outcomes $\pm 1$ appear with the probability $q^{(k)}_{\pm}$.} The corresponding CFIM of this measurement for every post-error-correction quantum state of Eq.~(\ref{eq:PEC})  is given by $F^{\rm PEC}_k$. Interestingly, this CFIM $F^{\rm PEC}_k$ is also singular showing the insufficiency of the measurement on the post-error-correction quantum states alone for the estimation of $\vec{B}$ (See the SM.). Indeed, it is only by combining the measurement outcomes of the stabilizers, described by probabilities $\{p_k\}$, with the measurement outcomes over the post-error-corrected states (\ref{eq:PEC}), described by probabilities $\{q_\pm\}$, that one can truly estimate the magnetic field $\vec{B}$. The overall precision is thus characterized by $\textrm{Tr}[F^{-1}]$, where the combined CFIM $F$ is given $F {=} F^{\rm stab}{+}F^{\rm PEC}$ and $F^{\rm PEC}{=}\sum_{k} p_k F^{\rm PEC}_k$ is the average post-error-correction CFIM. The overall achievable precision is bounded by  $\textrm{Tr}[F^{-1}]$  which scales asymptotically for large $N$ with shot-noise scaling $N^{-1}$ at best. Notice that while the measurement above is not optimal, even the ultimate precision $\textrm{Tr}[Q^{-1}]$, governed by the relevant QFIM $Q$ of the state $|\psi_{t}(\vec{B})\rangle$ without any error-correction, is ultimately shot-noise limited as well, albeit with a smaller magnitude. See appendix for details.  Nonetheless, a naive application of this protocol is inefficient for the following reason. \textcolor{black}{We note that Eq.~\eqref{eq:ecvm_rotation} indicates there is only partial suppression of the $X$-component of the effective logical qubit (GHZ probe) evolution, ultimately stemming from the fact that the $N$-qubit bit-flip code can only correct up to $\lfloor (N{-}1)/2 \rfloor$ flips. Thus, if we could fully suppress the $X$-component of the GHZ probe evolution by correcting all possible $N$ bit-flips, then we may improve the sensing performance. Remarkably, the $(N{+}1)$-qubit bit-flip code, where one of the qubits is shielded from the evolution induced by the magnetic field, achieves exactly this. This modified version of the protocol, schematically depicted in Fig.~\ref{fig:schematic}, is the main result of this letter, and will be described in the next section.}\\

\begin{table}[t]
\renewcommand{\arraystretch}{1.25}
\begin{tabular}{c||c|c|c}
\hline
\makecell{Source} & \multicolumn{3}{c}{Protocol with different probes} \\  \cline{2-4}
& \makecell{ancilla-free\\single probe} & \makecell{ancilla-assisted\\single-probe} & \makecell{ancilla-assisted\\dual-probe}
\\ \hline \hline
\makecell{Error\\suppression} & Partial along $X$ & Total along $X$ \quad  & Total along $X,Z$ \quad
\\ \hline \hline
\makecell{$F^\mathrm{stab}$} & singular & singular \quad  & Shot-noise \quad \\ \hline
\makecell{$F^\mathrm{PEC}$} & singular & singular & Heisenberg  \quad \\ \hline
\makecell{$F =$\\$F^{\textrm{stab}}{+}F^{\textrm{PEC}}$} & \makecell{At best shot-noise,\\ Does not saturate\\ $\mathrm{Tr}[Q^{-1}]$} & \makecell{Shot-noise, \\ saturates $\mathrm{Tr}[Q^{-1}]$}   &  \makecell{Heisenberg,  \\ saturates $\mathrm{Tr}[Q^{-1}]$} \\ \hline
\end{tabular}
\caption{Summary of sensing precision results for different variants of protocol described in the text.  $\mathrm{Tr}[Q^{-1}]$ computed with respect to $(B_x,B_z)$ for state $|\psi_{t}(\vec{B})\rangle$.}
\label{tab:summary}
\end{table}

\emph{Ancilla-assisted directional suppression.--} We now add a single ancilla qubit (topmost qubit in Fig.~\ref{fig:schematic}) to the original protocol. Thus, the probe now contains $N{+}1$ qubits. The goal is to completely suppress one of the components (say $X$) of the effective rotation of the logical qubit. Specifically, we initialize all $N{+}1$ qubits in the state $|\textrm{GHZ}_{Z}\rangle$. The magnetic field $\vec{B}$ then interacts for time $t$ with $N$ genuine probe qubits, but not with the single ancilla qubit. Following this, we measure the stabilizers and apply the corresponding correcting operation (see table~\ref{tab:end_matt_QECsyndromes} in appendix). Overall, the logical qubit basis is transformed as
\begin{align}
    & |0 \rangle_L \rightarrow |\tilde{0} \rangle_L \, \propto \, u_{00}^{N-k}u_{01}^k \, | 0 \rangle_L \nonumber \\
    & |1 \rangle_L \rightarrow|\tilde{1} \rangle_L  \, \propto \, u_{01}^ku_{11}^{N-k} \, |1 \rangle_L
\end{align}

\textcolor{black}{In contrast to Eq.~\eqref{eq:PEC}, the post-error-corrected states are now of the GHZ form $|0\rangle_L {+} e^{i\phi_k }|1\rangle_L$, with an acquired relative phase $\phi_k{=}2B_{\textrm{eff}}(N{-}k)$, which depends on an effective magnetic field $B_{\textrm{eff}}{=}\arctan[B_z\tan(Bt)/B]$.  In this case, the weight of the detected $X$-error, $k {\in} \left\{ \, 0, \, \ldots \,, \, N \, \right\}$, because the code can correct all possible errors as long as the first qubit is noiseless.  This is  explicitly shown for the $5$-qubit bit-flip code in Table~\ref{tab:end_matt_QECsyndromes} in appendix.  The relative phase $\phi_k$ acquired by the probe for each error weight $k$ can then be estimated optimally via usual single-parameter phase estimation. This optimal estimation, is in fact, obtained via  measuring the string operator $\Pi_k \hat{X}_k$ with binary measurement outcomes $\pm 1$. Now it is clear why we previously introduced this measurement setting to benchmark our results.} It superficially appears Heisenberg scaling is thus achieved (acquired relative phase $\phi_k {\propto} N$). However, the number of effective qubits in the probe is reduced by the number of detected errors $k$.  Therefore, the probe becomes progressively less useful with increasing $k$. In the extreme scenario, if $N$ errors are detected, then no relative phase is acquired. The combined CFIM $F{=}F^{\rm stab}+F^{\rm PEC}$ turns out to be invertible, and $\textrm{Tr}[F^{-1}]$ is expressed in the form  
\begin{equation}
   \textrm{Tr} [F^{-1}]  = \frac{1}{4Nt^2} \left( f_1^{(xx,z)} + f_1^{(zz,z)} + \frac{f_2^{(xx,z)} + f_2^{(zz,z)}}{f_3^{(z)} + N} \right),
\label{eq:cfim_single_ancilla_explicit}
\end{equation}
\begin{figure}
    \centering
    \includegraphics[width=\linewidth]{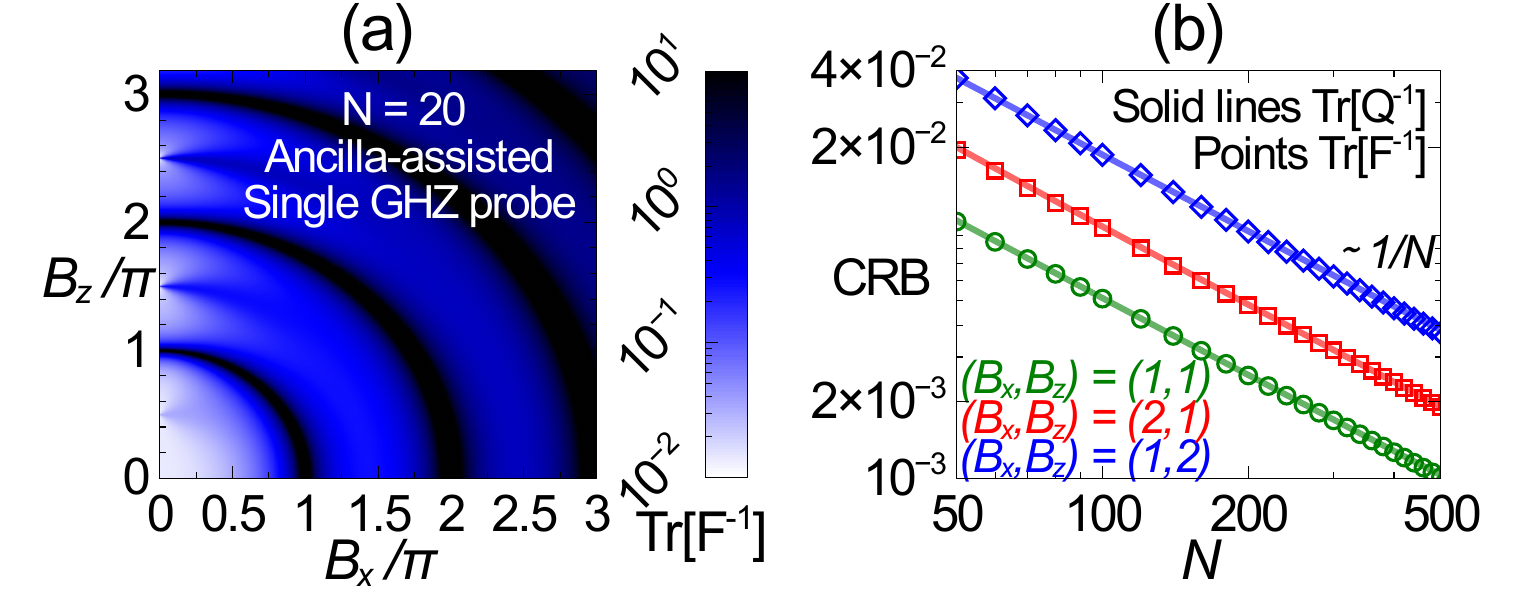}
    \caption{\textbf{Ancilla-assisted single GHZ probe :} (a) Heatmap of sensing precision quantified by $\mathrm{Tr}[F^{-1}]$ with respect to $(B_x,B_z)$. (b) Scaling with number of qubits $N$ of precision obtained via CFIM, i.e., $\mathrm{Tr}[F^{-1}]$ (hollow points) and via QFIM, i.e., $\mathrm{Tr}[Q^{-1}]$ (solid lines) for three different $(B_x,B_z)$ values ($t{=}1$ throughout).}
    \label{fig:single_probe}
\end{figure}
where $f_1^{(xx,z)}$, $f_2^{(xx,z)}$, $f_1^{(zz,z)}$, $f_2^{(zz,z)}$, $f_3^{(z)}$ are all non-negative functions independent of number of qubits $N$ (see appendix for explicit forms). We thus always obtain asymptotic shot-noise scaling for $\textrm{Tr}[F^{-1}]$ except when $ f_1^{(xx,z)} {=}  f_1^{(zz,z)} {=} 0$, which only occurs in trivial situations like $B_x{=}0$, i.e., when the magnetic field lies along the $Z$-direction with no $B_x$ to suppress, in which case we recover the well-known Heisenberg scaling for single-parameter estimation with GHZ probes \cite{giovannetti2011advances}. In Fig.~\ref{fig:single_probe}(a), we plot the heatmap of $\textrm{Tr}[F^{-1}]$ against magnetic fields $(B_x,B_z)$, which shows periodic behavior consistent with the magnetic field imprinting itself in the phase acquired by the probe.   In Fig.~\ref{fig:single_probe}(b), we demonstrate that the precision is generically shot-noise limited. Moreover, we show in Fig.~\ref{fig:single_probe}(b) that attainable precision for this strategy in fact coincides with the ultimate precision bound $\textrm{Tr}[Q^{-1}]$ quantified via the QFIM $Q$ of the output probe without any error correction. Notice the latter is only obtainable via a measurement setting generally dependent on parameter values $(B_x,B_z)$ themselves. In contrast, our protocol only calls for a fixed and separable measurement setting. Thus, the error-correction step in our protocol enables efficient extraction of information from a probe. This is the second main result of this letter. \\

\begin{figure}
    \centering
    \includegraphics[width=\linewidth]{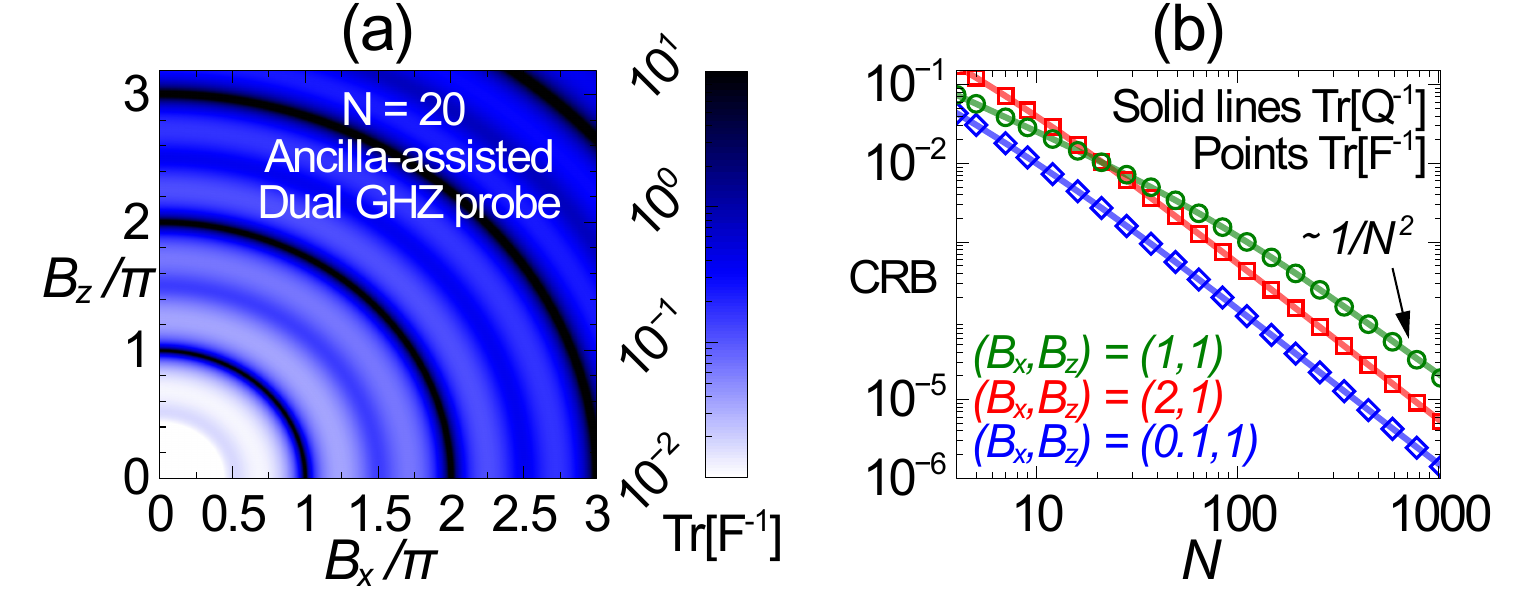}
    \caption{\textbf{Ancilla-assisted dual GHZ probe :} (a) Heatmap of sensing precision quantified by $\mathrm{Tr}[F^{-1}]$ with respect to $(B_x,B_z)$. (b) Scaling with number of qubits $N$ of precision obtained via CFIM, i.e., $\mathrm{Tr}[F^{-1}]$ (hollow points) and via QFIM, i.e., $\mathrm{Tr}[Q^{-1}]$ (solid lines) for three different $(B_x,B_z)$ values ($t{=}1$ throughout).}
    \label{fig:dual_probe}
\end{figure}

\emph{Recovery of Heisenberg scaling through dual probes.--} Having outlined the approach to directionally suppress one of the magnetic field directions, we now recall the precision still remained bounded by the shot-noise limit in the previous case. To recover Heisenberg scaling, we employ two $(N{+}1)$-qubit GHZ probes in tandem, i.e., with the probe qubits initialized in $|\textrm{GHZ}_Z\rangle\otimes|\textrm{GHZ}_X\rangle$, and two qubits being assigned as ancilla qubits. Concretely, the combined CFI matrix $F{=}F^{\rm stab}+F^{\rm PEC}$ is again invertible, and $\textrm{Tr}[F^{-1}]$ is expressed in the form 
\begin{equation}
    \textrm{Tr}[F^{-1}] = \frac{1}{4Nt^2} \left(\frac{1}{N+1} + \frac{2B^2t^2}{A + CN} \right), \label{eq_main_CFI}
\end{equation}
where $A{=}\sin^2(Bt) [3{-}\cos(2Bt)], C {=} \sin^2(Bt) [1{+}\cos(2Bt)]$, and is permutationally symmetric along $(B_x,B_z)$ as depicted in Fig.~\ref{fig:dual_probe}(a). From the expression above, $\textrm{Tr}[F^{-1}]$  is seen to exhibit Heisenberg scaling everywhere except when $C{=}0$, implying $Bt{=}m \pi {/}2$ for $m{\in}\mathbb{Z}$. Even integer values of $m$ correspond to full rotations, where our protocol fails because no relative phase is acquired. For odd integer values of $m$, while sensing is possible, the precision degrades to shot-noise scaling. For every other parameter value, genuine Heisenberg scaling is obtained in the asymptotic limit, as depicted in Fig.~\ref{fig:dual_probe}(b). Moreover, as demonstrated in Fig.~\ref{fig:dual_probe}(b), this strategy is again optimal, saturating the ultimate precision bound $\textrm{Tr}[Q^{-1}]$ expressed through QFIM $Q$ corresponding to the original probe $|\textrm{GHZ}_Z\rangle\otimes|\textrm{GHZ}_X\rangle$ having interacted with the magnetic field $\vec{B}$ without any further error correction. 

So far, we have quantified the sensing precision through CFIM, which is strictly valid only in the asymptotic regime. In order to demonstrate that our sensing protocol translates to genuine quantum advantage with limited number of runs $M$, one may employ Bayesian estimation to quantify its performance. As demonstrated in appendix, one can indeed reach Heisenberg scaling with respect to the probe size through Bayesian updating in the few-shot regime. \\

\emph{Conclusions.--} Quantum probes initialized in a GHZ state enable optimal  single-parameter quantum phase-interferometry via fixed separable measurements. Both quantum-enhancement of precision as well as the optimality of fixed separable measurements, disappear in the multiparameter regime. We formulate a protocol to harness QEC into GHZ-based multiparameter quantum sensing in which all but one parameter is treated as noise whose effects are corrected. Notably, while the GHZ probe is fundamentally shot-noise-limited in multi-parameter sensing, our protocol powered by QEC saturates the ultimate obtainable precision. This is achieved by introducing just a single shielded ancilla qubit into the GHZ state. Furthermore, we show that Heisenberg scaling is indeed recovered by using multiple complementary GHZ probes. We demonstrate the effectiveness of our protocol with concrete Bayesian inference simulations.  As QEC increasingly becomes experimentally viable \cite{422,Egan2021,DolevNeutral,google2025quantum,Quantinuum2024,LDPC_China}, similar protocols can be expected to augment multiparameter quantum metrology in many other sensing scenarios. 


\emph{Acknowledgments.--} AB acknowledges support from the  National Natural Science Foundation of China (grants No. 12274059, No. 12574528 and No. 1251101297). V.M. thanks support from the National Natural Science Foundation of China (grants No. 12374482 and No. W2432005).\\

\section{appendix}

In the appendix, we provide explicit formulations for certain expressions used in the main text and furnish a Bayesian estimation simulation of our protocol for completeness. \\

\emph{Error correction table.-- } As an illustration, we provide here the decoding lookup table for the $5$-qubit bit flip code under two variants of the protocol in this work. In the ancilla-free original formulation of the protocol, $X$ errors can occur on any qubit, and only weight-$1$ and weight-$2$ errors can be corrected.  An error of weight-$3$ and higher cannot be corrected because it gives rise to the same syndrome as a lower weight error obtained by multiplying the high-weight error by the logical $X_L$ operator of the code ($X_1X_2X_3X_4X_5$).  In general, every error $e$ of weight $k > \lfloor (N-1) / 2 \rfloor$ is incorrectly interpreted as an error $e' = e \, X_L$ of weight $k' \leq \lfloor (N-1) / 2 \rfloor$.  For example, $e=X_1X_2X_3X_4X_5$ is interpreted as $e'=I$, $e=X_1X_2X_3X_4$ is interpreted as $e'=X_5$, and so on.  In such cases, applying the correction to bring the state back to the codespace unavoidably causes a logical $X_L$ error.   However, for the ancilla-assisted protocol, no $X$ errors act on the first (ancilla) qubit, and all possible errors on the rest of the qubits can be corrected.  Notice that the corrections are the same for errors that do not have support on the first qubit.  For the other errors, the product of the two corrections is the logical $X_L$ operator.  This is generalizable to any $(N+1)$-qubit bit-flip code where $X$ errors only act on $N$ qubits.  Each one of the $2^N$ possible errors will give rise to a unique syndrome, since there are $N$ stabilizer generators and $2^N$ syndromes.  Table \ref{tab:end_matt_QECsyndromes} gives a list of the corresponding correction associated with each possible syndrome for the $5$-qubit bit-flip code.  


\begin{table}[h!]
\renewcommand{\arraystretch}{1}
\begin{tabular}{c|c|c|c||c|c}
\hline
\multicolumn{4}{c}{Stabilizer outcomes} & \multicolumn{2}{c}{Correction}  \\ \hline 
$Z_1Z_2$ & $Z_2Z_3$ & $Z_3Z_4$ & $Z_4Z_5$ & Ancilla-free & Ancilla-assisted \\ \hline
$+1$ & $+1$ & $+1$ & $+1$ & $I$ & $I$ \\ \hline
$+1$ & $+1$ & $+1$ & $-1$ & $X_5$ & $X_5$ \\ \hline
$+1$ & $+1$ & $-1$ & $+1$ & $X_4X_5$ & $X_4X_5$ \\ \hline
$+1$ & $+1$ & $-1$ & $-1$ & $X_4$ & $X_4$ \\ \hline
$+1$ & $-1$ & $+1$ & $+1$ & $X_1X_2$ & $X_3X_4X_5$ \\ \hline
$+1$ & $-1$ & $+1$ & $-1$ & $X_3X_4$ & $X_3X_4$ \\ \hline
$+1$ & $-1$ & $-1$ & $+1$ & $X_3$ & $X_3$ \\ \hline
$+1$ & $-1$ & $-1$ & $-1$ & $X_3X_5$ & $X_3X_5$ \\ \hline
$-1$ & $+1$ & $+1$ & $+1$ & $X_1$ & $X_2X_3X_4X_5$ \\ \hline
$-1$ & $+1$ & $+1$ & $-1$ & $X_1X_5$ & $X_2X_3X_4$ \\ \hline
$-1$ & $+1$ & $-1$ & $+1$ & $X_2X_3$ & $X_2X_3$ \\ \hline
$-1$ & $+1$ & $-1$ & $-1$ & $X_1X_4$ & $X_2X_3X_5$ \\ \hline
$-1$ & $-1$ & $+1$ & $+1$ & $X_2$ & $X_2$ \\ \hline
$-1$ & $-1$ & $+1$ & $-1$ & $X_2X_5$ & $X_2X_5$ \\ \hline
$-1$ & $-1$ & $-1$ & $+1$ & $X_1X_3$ & $X_2X_4X_5$ \\ \hline
$-1$ & $-1$ & $-1$ & $-1$ & $X_2X_4$ & $X_2X_4$ \\ \hline
\end{tabular}
\caption{Decoding lookup table for the $5$-qubit bit-flip code (probe prepared in a $|\textrm{GHZ}_\textrm{Z}\rangle$ state).  In the ancilla-free case, $X$ errors act on all $5$ qubits and only weight-$1$ and weight-$2$ errors can be corrected.  In the ancilla-assisted case, the first qubit is noiseless so we can modify the decoding strategy and correct all possible errors on the other $4$ qubits.  This strategy is generalizable since a $(N+1)$-qubit bit-flip code has $2^N$ different syndromes and $2^N$ different possible $X$ errors on the last $N$ qubits.  Each error gives rise to a unique syndrome and all errors can be unambiguously detected and corrected.}
\label{tab:end_matt_QECsyndromes}
\end{table}

\begin{figure}
    \centering
    \includegraphics[width=0.8\linewidth]{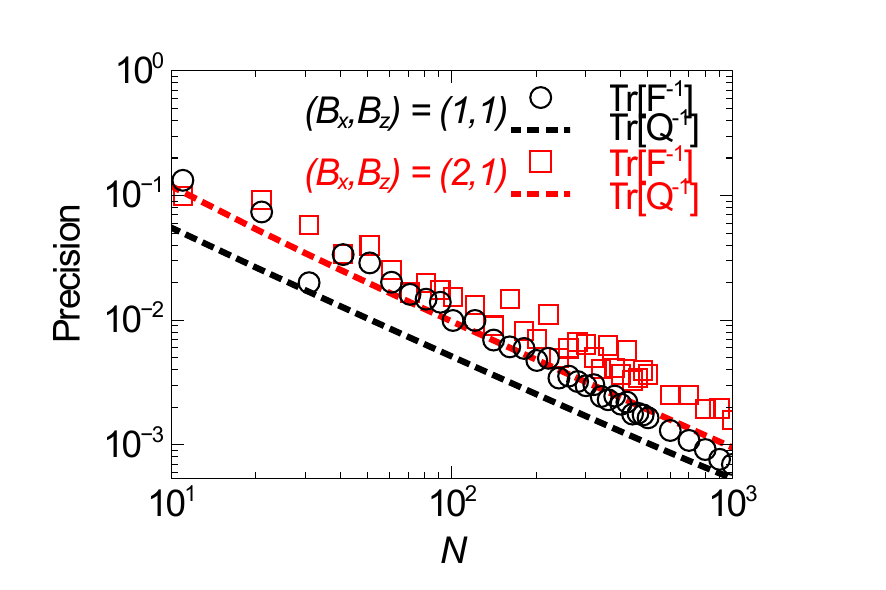}
    \caption{\textbf{Ancilla-free single GHZ probe :} Precision Scaling with number of qubits $N$ of precision obtained via CFIM, i.e., $\mathrm{Tr}[F^{-1}]$ (hollow points) and via QFIM, i.e., $\mathrm{Tr}[Q^{-1}]$ (solid lines) for two different $(B_x,B_z)$ values ($t{=}1$ throughout).}
    \label{fig:ancilla_free}
\end{figure}

\emph{Complete expressions of post-error-corrected states.-- } For the ancilla-free original protocol, the post-error-corrected state $|\psi_{k}^{\textrm{PEC}}\rangle $ is given up to a normalization factor by 
\begin{align}
    |\psi_{k}^{\textrm{PEC}}\rangle\sim |0\rangle^{\otimes N} + \frac{B_x^{N-2k} + (-B_z + iB \cot Bt)^{N-2k}}{B_x^{N-2k} {+} (B_z {+} iB \cot Bt)^{N-2k} \left(\frac{B_z+iB\cot Bt}{-B_z + iB\cot Bt}\right)^k}|1\rangle^{\otimes N}  
\end{align}

\noindent For the ancilla-assisted single GHZ probe initialized along the $Z$-basis, the post-error-corrected state $|\psi_{k}^{\textrm{PEC}}\rangle $ is given by 
\begin{align}
     |\psi_{k}^{\textrm{PEC}}\rangle = \frac{|0\rangle^{\otimes N} + e^{2i(N-k)\arctan[\frac{B_z \tan Bt}{B}]} |1\rangle^{\otimes N} }{\sqrt{2}}
\end{align}

\noindent For the ancilla-assisted single GHZ probe initialized along the $X$-basis, the post-error-corrected state $|\psi_{k}^{\textrm{PEC}}\rangle$ is given by 
\begin{align}
     |\psi_{k}^{\textrm{PEC}}\rangle = \frac{|+\rangle^{\otimes N} + e^{2i(N-k)\arctan[\frac{B_x \tan Bt}{B}]} |-\rangle^{\otimes N} }{\sqrt{2}}
\end{align}

\emph{Outcome probabilities for ancilla-free protocol.--} For $k$-errors, outcome probabilities $p_k$ are given by 
\begin{eqnarray}
  p_k & =  \binom{N}{k} [( \cos^2(Bt) + n_z^2 \sin^2(Bt))^{N-k} (n_x^2 \sin^2(Bt))^{k} \nonumber \\
 & + ( \cos^2(Bt) + n_z^2 \sin^2(Bt) )^{k} ( n_x^2 \sin^2(Bt))^{N-k}],
\end{eqnarray} 
\noindent where $k {\in} \mathbb{Z}~\textrm{and}~ k{\in} [0, \lfloor \frac{N-1}{2}\rfloor]$. The form reflects the symmetric repetition of syndromes when $k{>} \frac{\lfloor N-1\rfloor}{2}$ as described earlier. In Fig.~\ref{fig:ancilla_free}, we use this expression to depict $\mathrm{Tr}[F^{-1}]$ with number of probe qubits $N$, and show that the ancilla-free protocol does not saturate the ultimate precision bound governed by $\mathrm{Tr}[Q^{-1}]$, with $Q$ being the QFIM of the output state $|\psi_t(\vec{B})\rangle$ introduced in the main text. \\

\begin{figure}
    \centering
    \includegraphics[width=0.9\linewidth]{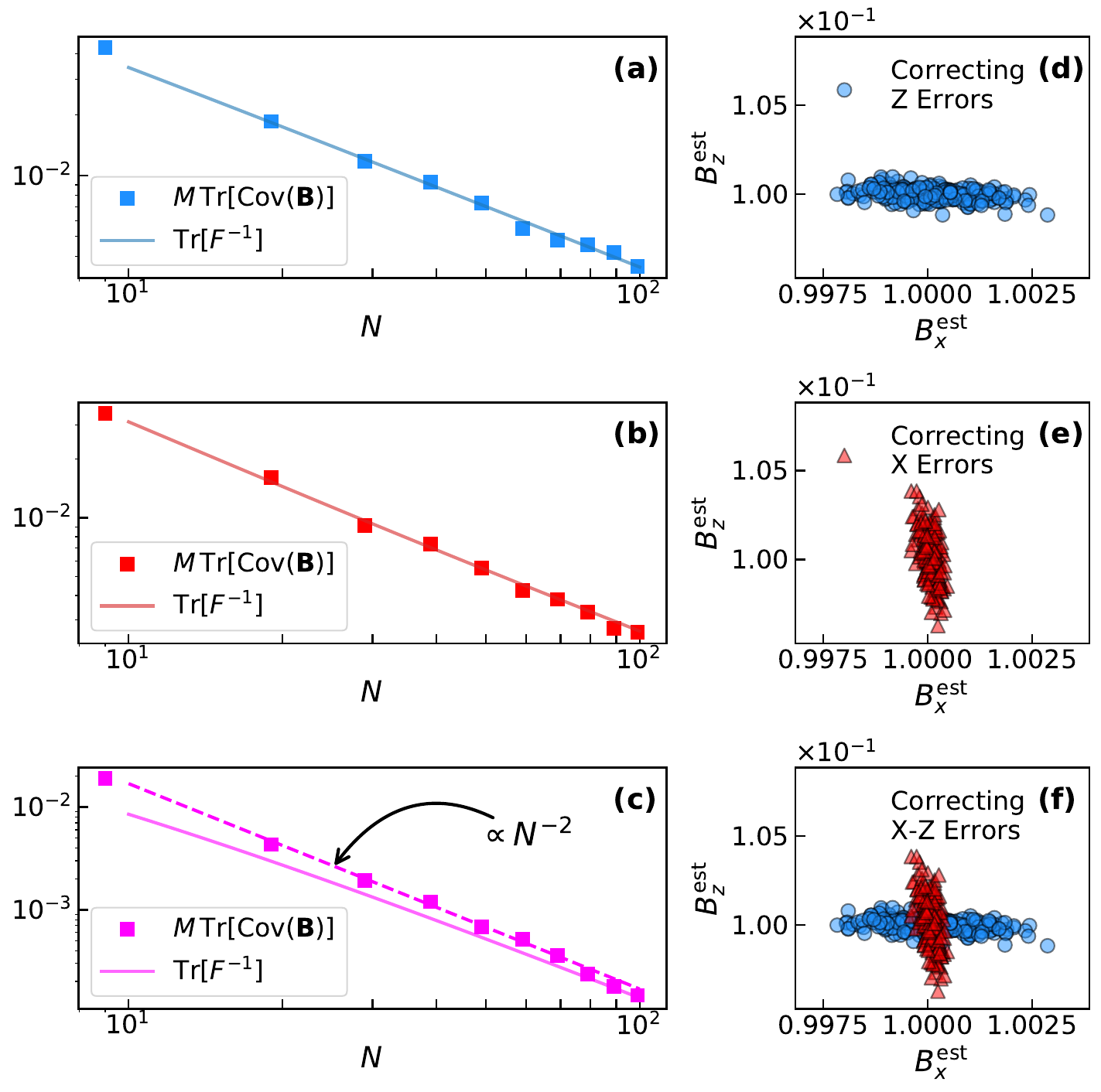}
    \caption{\textbf{Heisenberg limit with Bayesian analysis:} The left column shows the trace of the rescaled covariance matrix $M\mathrm{Cov}[\mathfrak{B}]$ alongside the inverse trace of the CFIM $\mathrm{Tr}[F^{-1}]$ as functions of $N$. Panel (a) corresponds to a single probe state $|\text{GHZ}_X\rangle$ that corrects $Z$ errors, panel (b) to a single probe $|\text{GHZ}_Z\rangle$ correcting $X$ errors, and panel (c) to the combined use of two probes, $|\text{GHZ}_X\rangle$ and $|\text{GHZ}_Z\rangle$, correcting $X$ and $Z$ errors, respectively. The right column shows the estimated values $(B_x^{\mathrm{est}}, B_z^{\mathrm{est}})$ for $N = 50$. Each point represents an estimate obtained from $M = 4000$ measurements, using either the single probe $|\textrm{GHZ}_X\rangle$, $|\textrm{GHZ}_Z\rangle$, or both.}
    \label{fig_bayesian}
\end{figure} 

\emph{Outcome probabilities for ancilla-assisted protocol.--} For $k$-errors, the acquired phase angle is given by $\phi_k {=}2 (N-k) \arctan[B_z \tan(Bt)/B]$, and outcome probabilities $p_k$ are given by 
\begin{equation}
  p_k =  \binom{N}{k} \left( \cos^2(Bt) + n_z^2 \sin^2(Bt) \right)^{N-k} \left( n_x^2 \sin^2(Bt) \right)^{k}. 
\end{equation} 
\emph{Complete expressions of CFIM elements of ancilla-assisted protocol. --} The individual functions constituting the CFIM quoted in the main text for the ancilla-assisted protocol are given as 
\begin{align}
    f_1^{(xx,z)} & {=} \frac{[2n_z^2Bt + n_x^2\sin(2Bt)]^2}{4\sin^2(Bt)[1{-}n_x^2\sin^2(Bt)]}; f_2^{(xx,z)}{=} \frac{n_x^2n_z^2[1{-}Bt\cot(Bt)]^2}{1-n_x^2\sin^2(Bt)} ; \\
    f_1^{(zz,z)} &{=} \frac{n_x^2n_z^2[2Bt -\sin(2Bt)]^2}{4\sin^2(Bt)[1-n_x^2\sin^2(Bt)]}; f_2^{(zz,z)}{=}  \frac{[n_z^2+n_x^2Bt\cot(Bt)]^2}{1-n_x^2\sin^2(Bt)} ;\\
    f_3^{(z)} &{=} \frac{n_x^2\sin^2(Bt)}{1-n_x^2\sin^2(Bt)} ;  \quad \textrm{where} ~ n_x{=}B_x/B, n_z{=}B_z/B.
\end{align}

\textit{Bayesian sensing protocol---} To explicitly demonstrate that the Heisenberg limit of precision in Eq.~\eqref{eq_main_CFI} is practically attainable, we simulate the estimation of two unknown magnetic field components, $B_x$ and $B_z$, using a Bayesian inference approach. Bayesian inference is a widely used and powerful estimation method that systematically updates prior knowledge about unknown parameters based on observed measurement data. The complete Bayesian procedure, along with technical details, is provided in the SM. Figs.~\ref{fig_bayesian}(a)–(c) show the trace of the rescaled covariance matrix $M\mathrm{Cov}[\mathfrak{B}]$ and the inverse of the CFIM as functions of $N$. We consider three cases: (a) a single probe correcting $Z$ errors, (b) a single probe correcting $X$ errors, and (c) two probes correcting $X$ and $Z$ errors, respectively. As shown in Figs.~\ref{fig_bayesian}(a)–(c), the Bayesian analysis---relying only on the observed measurement data---clearly reaches the theoretical precision bounds set by corresponding $\mathrm{Tr}[F^{-1}]$ expressions in the main text vide Eq.~\eqref{eq:cfim_single_ancilla_explicit} and Eq.~\eqref{eq_main_CFI}. In particular, Fig.~\ref{fig_bayesian}(c) clearly demonstrates that using two probes, each correcting for $X$ and $Z$ errors respectively, achieves the Heisenberg limit of precision. This is evidenced by the $N^{-2}$ scaling trend observed in the covariance matrix. To further clarify how the theoretical bounds are attained, Figs.~\ref{fig_bayesian}(d)–(f) illustrate the estimated values of $B_x^{\mathrm{est}}$ and $B_z^{\mathrm{est}}$ for $N = 50$. Each point $(B_x^{\mathrm{est}}, B_z^{\mathrm{est}})$ represents an estimate obtained from $M = 4000$ measurements using a single probe. We simulated 100 independent experiments, from which the covariance matrix was computed. As seen in Figs.~\ref{fig_bayesian}(d)–(f), each single probe reduces the uncertainty in the magnetic field component it is designed to correct. Intuitively, combining two probes---as shown in Fig.~\ref{fig_bayesian}(f)---allows us to discard inconsistent estimated tuples and reinforce the estimates that overlap within a common region. This overlap effectively narrows the range of plausible values, improving overall estimation precision. \\

\bibliographystyle{apsrev4-2}
\bibliography{qec_qsense}

\newpage
\onecolumngrid

 \vspace{1em}
 \begin{center}
   {\large \bfseries Supplemental Material : Quantum error-correction for multi-parameter metrology}\\[8pt]
 \end{center}
 \vspace{1em}
In this supplemental material, we provide some detailed explanations and analysis for results quoted in the main text. 

\section{Benchmarking without quantum error correction}
Our probe will be a system of $N$ qubits.  We initialize it in the GHZ state $ \left (|0 \rangle ^{\otimes N} + |1 \rangle ^{\otimes N} \right) / \sqrt{2} $, which is optimal if the magnetic field is aligned in the $Z$ direction \cite{giovannetti2011advances}.  The unknown magnetic field in the $XZ$ plane generates the following Hamiltonian on the probe: 
\begin{equation}
    \hat{H} = B_x \sum_{j=0}^{N} \hat{X}_{j} + B_z \sum_{j=0}^{N} \hat{Z}_{j}, \label{eq_H_probe}
\end{equation}
where $B_x$ and $B_z$ are the components of the magnetic field in the $X$ and $Z$ direction, respectively, and $\hat{X}_{j}$ and $\hat{Z}_{j}$ are the Pauli $X$ and $Z$ matrices acting on qubit $j$, respectively.  Setting $\hbar = 1$ and $t=1$, the time evolution operator is then given by the following rotation:

\begin{align}
    \hat{\mathfrak{U}} &= \exp(-i \hat{H}t) \\ 
         &= [\exp(-it (B_x \hat{X} + B_z \hat{Z})) ] ^{\otimes N} \\
         &= [\cos(Bt) - i\sin(Bt) (n_x \hat{X} + n_z \hat{Z}) ] ^{\otimes N} \\
         &= \begin{pmatrix}
            \cos(Bt) - i n_z \sin(Bt) & - i n_x \sin(Bt) \\
            - i n_x \sin(Bt) & \cos(B) + i n_z \sin(Bt)
\end{pmatrix}^{\otimes N} \label{eq:matrix}
\end{align}

where $B = \sqrt{B_x^2 + B_z^2}$, $n_x = B_x/B$, $n_z = B_z / B$.

\subsection{Ultimate theoretical limit of precision}

After interacting with the magnetic field, the probe (now in state $| \psi \rangle = \hat{\mathfrak{U}} | \textrm{GHZ} \rangle$) acquires information about the unknown values of $B_x$ and $B_z$.  The ultimate precision limit for the joint estimation of these two parameters is given by the quantum Fisher information matrix (QFIM), whose elements correspond to:

\begin{equation}
    [Q]_{ab} = 4 \, \textrm{Re} \left[ \, \langle \partial_a \psi | \partial_b \psi \rangle - \langle \partial_a \psi | \psi \rangle \langle \psi |\partial_b \psi \rangle \, \right],
\end{equation}
where $\partial_a$ denotes the partial derivative with respect to parameter $a$, and $a$ and $b$ refer to the unknown parameters to be estimated ($B_x$ and $B_z$).  For a probe prepared in the state $|\textrm{GHZ}_Z \rangle$, the QFIM elements simplify to:
\begin{equation}
[Q]_{ab} = 2 \sum_{k=0}^N \binom{N}{k} \partial_a(\mathfrak{U}_{N,k}^*) \partial_b (\mathfrak{U}_{N,k}) \, - \, \left( \sum_{k=0}^N \binom{N}{k}  \partial_a(\mathfrak{U}_{N,k}^*) \, \mathfrak{U}_{N,k} \right) \left( \sum_{k=0}^N \binom{N}{k} \mathfrak{U}_{N,k}^* \, \partial_b(\mathfrak{U}_{N,k}) \right),
\end{equation}
where $^*$ denotes the complex conjugate and
\begin{align}
    \mathfrak{U}_{N,k} &= u_{00}^{N-k}u_{01}^k + u_{01}^{N-k} u_{11}^k \\
    u_{00}  &= \cos(Bt) - in_z\sin(Bt) \label{eq:R00} \\
    u_{01}  &= -in_x\sin(Bt) \label{eq:R01} \\
    u_{11}  &= \cos(Bt) + in_z \sin(Bt) \label{eq:R11} .
\end{align}

Notice that $u_{00}$, $u_{01}$, and $u_{11}$ correspond to the matrix elements of the single-qubit operator defined in equation \eqref{eq:matrix}.  On the other hand, for a probe prepared in the state $|\textrm{GHZ}_X \rangle$, the QFIM elements simplify to:
\begin{equation}
[Q]_{ab} = 2 \sum_{k=0}^N \binom{N}{k} \partial_a(\tilde{\mathfrak{U}}_{N,k}^*) \partial_b (\tilde{\mathfrak{U}}_{N,k}) \, - \, \left( \sum_{k=0}^N \binom{N}{k}  \partial_a(\tilde{\mathfrak{U}}_{N,k}^*) \, \tilde{\mathfrak{U}}_{N,k} \right) \left( \sum_{k=0}^N \binom{N}{k} \tilde{\mathfrak{U}}_{N,k}^* \, \partial_b(\tilde{\mathfrak{U}}_{N,k}) \right),
\end{equation}
where
\begin{align}
    \tilde{\mathfrak{U}}_{N,k} &= u_{++}^{N-k}u_{+-}^k + u_{+-}^{N-k} u_{--}^k \\
    u_{++}  &= \cos(Bt) - in_x\sin(Bt) \label{eq:R++} \\
    u_{+-}  &= -in_z\sin(Bt) \label{eq:R+-} \\
    u_{--}  &= \cos(Bt) + in_x \sin(Bt) \label{eq:R--} .
\end{align}

Notice that $u_{++}$, $u_{+-}$, and $u_{--}$ correspond to the matrix elements of the same single-qubit operator defined in equation \eqref{eq:matrix}, but in the $X$ basis.  Neither of these two QFI matrices allow us to reach Heisenberg scaling.


If we use $2N$ qubits and prepare the probe in the state $|\textrm{GHZ}_Z \rangle \otimes |\textrm{GHZ}_X \rangle$, the resulting QFIM elements of the joint probe simplify to:
\begin{equation}
\begin{split}
[Q]_{ab} &= \sum_{k_x=0}^N \sum_{k_z=0}^N \binom{N}{k_x} \binom{N}{k_z} \partial_a(\mathfrak{U}_{N,k_1}^* \tilde{\mathfrak{U}}_{N,k_2}^*) \partial_b (\mathfrak{U}_{N,k_1} \tilde{\mathfrak{U}}_{N,k_2}) \\
               & - \frac{1}{4} \, \left( \sum_{k_x=0}^N \sum_{k_z=0}^N \binom{N}{k_x} \binom{N}{k_z}  \partial_a(\mathfrak{U}_{N,k_1}^* \tilde{\mathfrak{U}}_{N,k_2}^*) \, \mathfrak{U}_{N,k_1} \tilde{\mathfrak{U}}_{N,k_2} \right) \left( \sum_{k_x=0}^N \sum_{k_z=0}^N \binom{N}{k_x} \binom{N}{k_z} \mathfrak{U}_{N,k_1}^* \tilde{\mathfrak{U}}_{N,k_2}^* \, \partial_b(\mathfrak{U}_{N,k_1} \tilde{\mathfrak{U}}_{N,k_2}) \right).
\end{split}
\end{equation}

We may simplify this expression by noting  $Q[|\textrm{GHZ}_Z\rangle\otimes|\textrm{GHZ}_X\rangle]{=}Q[|\textrm{GHZ}_Z\rangle] + Q[|\textrm{GHZ}_X\rangle]$, and replacing the individual QFI elements for N-qubit QFIMs as stated before.  This QFIM results in Heisenberg scaling. For the state of the probe after interacting with the magnetic field, $|\psi \rangle {=} (\hat{\mathfrak{U}} {\otimes} \hat{\tilde{\mathfrak{U}}})( |\textrm{GHZ}_Z\rangle {\otimes} |\textrm{GHZ}_X\rangle)$, the multi-parameter quantum Cram\'er-Rao bound is saturable, since $\textrm{Im}(|\langle \partial_a \psi| \partial_b \psi \rangle) = 0, \, \, \forall \, a, b$ (See Theorem $3.2$ of \cite{Liu2020}).  This implies that there exists a measurement basis set whose classical Fisher information matrix (CFIM) will be equal to the QFIM.  However, in general, this measurement scheme will depend on the parameters $B_x$ and $B_z$, which are precisely the unknown values that we are trying to estimate.  Therefore, in practice, saturating the quantum Cram\'er-Rao bound is not achievable in most cases involving multi-parameter sensing.



\section{Ancilla-assisted protocol with single GHZ-probe}

Our  probe will be a system of $N+1$ qubits.  We first prepare the probe in the state $| \textrm{GHZ}_Z \rangle = (1/\sqrt{2}) \left (|0 \rangle ^{\otimes (N+1)} + |1 \rangle ^{\otimes (N+1)} \right)$, which is known to be optimal if the magnetic field is aligned in the $Z$ direction. We let the system evolve for a time $t$.  Crucially, during this time evolution, the magnetic field will act only on $N$ qubits.  This will allow us to completely suppress the $X$-component of the field.  The Hamiltonian of our system is:

\begin{equation}
    H = B_x \sum_{j=0}^{N} \sigma_x^{(j)} + B_z \sum_{j=0}^{N} \sigma_z^{(j)},
\end{equation}
where $\sigma_x^{(j)}$ and $\sigma_z^{(j)}$ are the Pauli X and Z matrices acting on qubit $j$, respectively. Remarkably, since the Hamiltonian only acts on $N$ qubits and we are using the $(N+1)$-qubit bit-flip code we can correct all possible $X$ errors acting on $N$ qubits. By measuring the stabilizers generators and performing error correction, we can completely suppress the $X$ component of the magnetic field. 
After error correction, the effective time evolution will be rotation of the original GHZ about the $Z$ axis.  The final state can be expressed as:
\begin{equation}
    |\psi \rangle = \frac{1}{\sqrt{2}} \left(|0 \rangle ^{\otimes (N+1)} + e^{i \phi_k} |1 \rangle ^{\otimes (N+1)} \right) 
\end{equation}

The acquired relative phase angle $\phi_k$ will depend on the subspace onto which the state was projected during the measurement of the stabilizers.  This outcome is random, but (1) we know which subspace we're at (since we know the classical outcome of the stabilizers) and (2) we know the probabilities of each possible outcome.The probabilities of each outcome and the acquired relative phase only depend on the number of detected errors, not on the specific locations of these errors.  Table \ref{tab:2D} summarizes the results. The effective magnetic field after error correction is given by:
\begin{equation}
    B_{\textrm{eff}} = \arctan[n_z \tan(Bt)].
\end{equation} Notice that when the magnetic field points exactly in the $Z$ direction ($n_z {=} 1$), the probability of the first outcome is $1$, the effective magnetic field $B_{\textrm{eff}} {=} B{=} B_z $, and the acquired relative phase is exactly what one expects for an optimal probe of $N$ qubits.

\vspace{20pt}
\begin{table}[ht]
\centering
\renewcommand{\arraystretch}{1.5} 
\begin{tabular}{c|c|c}
\hline
Number of $X$ errors detected & Probability of occurrence ($p_k$) & Relative phase angle ($\phi_k$)  \\ \hline
$0$ & $(\cos^2(Bt) + n_z^2 \sin^2(Bt))^{N}$ & $2 B_{\textrm{eff}} N$ \\
$1$ & $\binom{N}{1} \left( \cos^2(Bt) + n_z^2 \sin^2(Bt) \right)^{N-1} \left( n_x^2 \sin^2(Bt) \right)$ & $2 B_{\textrm{eff}} (N-1)$ \\ 
$\vdots$ & $\vdots$  & $\vdots$ \\
$k$ & $\binom{N}{k} \left( \cos^2(Bt) + n_z^2 \sin^2(Bt) \right)^{N-k} \left( n_x^2 \sin^2(Bt) \right)^{k}$ & $2 B_{\textrm{eff}} (N-k)$ \\
$\vdots$ & $\vdots$  & $\vdots$ \\
$N$ & $\left( n_x^2 \sin^2(Bt) \right) ^{N}$ & 0 \\ \hline
\end{tabular}
\caption{Probabilities of detecting $k$ ``errors'' and resulting relative phases after error correction for an ($N+1$)-qubit $|\textrm{GHZ}\rangle_Z$ probe and a magnetic field acting on all qubits except the first one.  The magnetic field $\vec{B} = (B_x,B_z)$ is in the $XZ$ plane, $n_x=B_x/B$, $n_z=B_z/B$, and $B=|\vec{B}|=\sqrt{B_x^2+B_z^2}$.}
\label{tab:2D}
\end{table}

\subsection{Ancilla-assisted protocol with single GHZ-probe in the X basis}

We can instead initialize our system in the state $| \textrm{GHZ}_X \rangle = (1/\sqrt{2}) \left (|+ \rangle ^{\otimes (N+1)} + |- \rangle ^{\otimes (N+1)} \right)$, where $|\pm \rangle = (1/\sqrt{2}) (|0 \rangle \pm |1 \rangle)$, which can be regarded as the logical $|+ \rangle_L$ of the $(N+1)$-qubit phase-flip code.  After letting all the qubits except the first one interact with the magnetic field $\vec{B}$ for a time $t$ and measuring the stabilizer generators $\{ X_1X_2, X_2X_3, \ldots, X_NX_{N+1} \}$ and doing error correction, the effective time evolution will be rotation of the original GHZ about the $X$ axis.  The final state can be expressed as:
\begin{equation}
    |\psi \rangle = \frac{1}{\sqrt{2}} \left(|+ \rangle ^{\otimes (N+1)} + e^{i \phi_{k_z}} |- \rangle ^{\otimes (N+1)} \right) 
\end{equation}

The acquired relative phase angle $\phi_{k_z}$ will again depend on the subspace onto which the state was projected during the measurement of the stabilizers.  Notice that we use the subscript $k_z$ to denote the number of $Z$ errors detected.  Table \ref{tab:2D-Xbasis} summarizes the results.

\vspace{20pt}
\begin{table}[ht]
\centering
\renewcommand{\arraystretch}{1.5} 
\begin{tabular}{c|c|c}
\hline
Number of $Z$ errors detected & Probability of occurrence ($p_{k_z}$) & Relative phase angle ($\phi_{k_z}$)  \\ \hline
$0$ & $(\cos^2(Bt) + n_x^2 \sin^2(Bt))^{N}$ & $2 B_{\textrm{eff}}^{(X)} N$ \\
$1$ & $\binom{N}{1} \left( \cos^2(Bt) + n_x^2 \sin^2(Bt) \right)^{N-1} \left( n_z^2 \sin^2(Bt) \right)$ & $2 B_{\textrm{eff}}^{(X)} (N-1)$ \\ 
$\vdots$ & $\vdots$  & $\vdots$ \\
$k$ & $\binom{N}{k} \left( \cos^2(Bt) + n_x^2 \sin^2(Bt) \right)^{N-k} \left( n_z^2 \sin^2(Bt) \right)^{k}$ & $2 B_{\textrm{eff}}^{(X)} (N-k)$ \\
$\vdots$ & $\vdots$  & $\vdots$ \\
$N$ & $\left( n_z^2 \sin^2(Bt) \right) ^{N}$ & 0 \\ \hline
\end{tabular}
\caption{Probabilities of detecting $k$ ``errors'' and resulting relative phases after error correction for an ($N+1$)-qubit $|\textrm{GHZ}_X\rangle$ probe and a magnetic field acting on all qubits except the first one.  The magnetic field $\vec{B} = (B_x,B_z)$ is in the $XZ$ plane, $n_x=B_x/B$, $n_z=B_z/B$, and $B=|\vec{B}|=\sqrt{B_x^2+B_z^2}$.}
\label{tab:2D-Xbasis}
\end{table}
\vspace{20pt}

The effective magnetic field after error correction is given by:
\begin{equation}
    B_{\textrm{eff}}^{(X)} = \arctan[n_x \tan(Bt)].
\end{equation}

\subsection{Complete suppression of $X$-errors}

The key to completely suppress the $X$ component of the magnetic field lies in using a bit-flip code capable of correcting all possible $X$ errors.  For an $(N+1)$-qubit bit-flip code, if $X$ errors can act on every possible qubit, then only errors of at most weight-$n$ can be corrected, where $n = \lfloor N/2 \rfloor$. For $N$ even, this means that exactly half of the possible errors are uncorrectable.  Since the continuous rotation generated by the field will be randomly discretized to any of these possible errors, the uncorrectable errors will cause the $X$ component of the field to not be completely suppressed. However, if $X$ errors can act on all but one of the qubits, and we know which one is the noiseless qubit, it is possible to correct all possible errors.

\subsection{Analysis of scaling of precision}

There are several sources of information that we can use to estimate $B_x$ and $B_z$.  In the first place, we can use the number of detected X ``errors'', $k$, which derives directly from the stabilizer outcomes.  The corresponding $2 \times 2$ CFI matrix corresponds to:
\begin{equation}
    [F^{\textrm{stab}}]_{ab} = \sum_{k=0}^{N} \frac{1}{p_k} \partial_a (p_k) \partial_b (p_k),
\end{equation}
where the subscripts $a$ and $b$ refer to the parameters to be estimated: $B_x$, $B_z$. This matrix is singular for every value of $N$, which implies that we cannot estimate both $B_x$ and $B_z$. Secondly, we can also perform single-parameter estimation with the final state after QEC.   

\begin{equation}
    [F^{\textrm{PEC}}]_{ab} = \sum_{k=0}^{N} p_k [F^{\textrm{PEC}}_{k}]_{ab}
\end{equation}
where the subscripts $a$ and $b$ refer to the parameters to be estimated: $B_x$, $B_z$.

\begin{equation}
[F^{\textrm{PEC}}_{k}]_{ab} = \frac{1}{q_{k,+}} \partial_a (q_{k,+}) \partial_b (q_{k,+}) + \frac{1}{q_{k,-}} \partial_a (q_{k,-}) \partial_b (q_{k,-}), \quad q_{k,+} = \cos^2 \left( B_{\textrm{eff}}(N-k) \right), q_{k,-} = \sin^2 \left( B_{\textrm{eff}}(N-k) \right)
\end{equation}

Notice that $[F^{\textrm{PEC}}]_{ab}$ is the average of all the individual $[F^{\textrm{PEC}}_{k}]_{ab}$ (CFI for the single-parameter sensing after $k$ errors were detected) weighted by the probability of detecting $k$ errors.  The sum is carried up to $k=N-1$ instead of $k=N$, because when $N$ errors are detected the final state after QEC does not pick any relative phase and no information can be extracted. Finally, the total CFIM that takes into account all possible stabilizer measurement outcomes as well as the single-parameter sensing done after QEC is given by:

\begin{align*}
    [F_{\textrm{total (1 GHZ$_Z$)}}]_{ab} &= \sum_{k=0}^{N} \left( \frac{1}{p_k \, q_{k,+}} \partial_a (p_k \, q_{k,+}) \partial_b (p_k \, q_{k,+}) + \frac{1}{p_k \, q_{k,-}} \partial_a (p_k \, q_{k,-}) \partial_b (p_k \, q_{k,-}) \right) \\
                                      &=  \underbrace{\sum_{k=0}^{N} \frac{1}{p_k} \partial_a (p_k) \partial_b (p_k)}_{[F^{\textrm{stab}}]_{ab}} + \underbrace{\sum_{k=0}^{N-1} p_k \left( \frac{1}{q_{k,+}} \partial_a (q_{k,+}) \partial_b (q_{k,+}) + \frac{1}{q_{k,-}} \partial_a (q_{k,-}) \partial_b (q_{k,-}) \right)}_{[F^{\textrm{PEC}}]_{ab}} + \underbrace{\textrm{cross terms}}_{0} \\
                                      &= [F^{\textrm{stab}}]_{ab} + [F^{\textrm{PEC}}]_{ab}
\end{align*}

Notice that this is just the sum of the individual CFIs from the stabilizer outcomes and the single-parameter estimation after QEC, since the cross terms are equal to $0$. If we start with a GHZ state in the $X$ basis ($(|+\rangle ^{\otimes N} + |-\rangle ^{\otimes N})/\sqrt{2}$), and then measure the stabilizers of the phase-flip code to correct the $Z$ errors, we get similar expressions.  The total CFI matrix using both probes is given by:  
\begin{equation}
    F_{\textrm{total (2 GHZ's)}} = F_{\textrm{total (1 GHZ$_Z$)}} + F_{\textrm{total (1 GHZ$_X$)}}
\end{equation}

We can obtain analytical expressions for every CFI matrix.  If we only use a probe in the $Z$ basis $|\textrm{GHZ}_Z$, after inverting the CFIM, we get the following expressions for the diagonal matrix elements of the inverse matrix:
\begin{equation}
    [F_{\textrm{total (GHZ$_Z$)}}^{-1}]_{B_x B_x} = \frac{1}{4Nt^2} \left( f_1^{(xx,z)} + \frac{f_2^{(xx,z)}}{f_3^{(z)} + N} \right), \, \quad  [F_{\textrm{total (GHZ$_Z$)}}^{-1}]_{B_z B_z} = \frac{1}{4Nt^2} \left( f_1^{(zz,z)} + \frac{f_2^{(zz,z)}}{f_3^{(z)} + N} \right) \,
\end{equation}


\begin{equation}
    \textrm{where} \quad f_1^{(xx,z)} = \frac{[2n_z^2Bt + n_x^2\sin(2Bt)]^2}{4\sin^2(Bt)[1-n_x^2\sin^2(Bt)]} \, \, , \, \, f_2^{(xx,z)} = \frac{n_x^2n_z^2[1-Bt\cot(Bt)]^2}{1-n_x^2\sin^2(Bt)} 
\end{equation}

\begin{equation}
    f_1^{(zz,z)} = \frac{n_x^2n_z^2[2Bt -\sin(2Bt)]^2}{4\sin^2(Bt)[1-n_x^2\sin^2(Bt)]} \, \, , \, \, f_2^{(zz,z)} = \frac{[n_z^2+n_x^2Bt\cot(Bt)]^2}{1-n_x^2\sin^2(Bt)}, \quad f_3^{(z)} = \frac{n_x^2\sin^2(Bt)}{1-n_x^2\sin^2(Bt)}, 
\end{equation} 

Notice that $[F_{\textrm{total (GHZ$_Z$)}}^{-1}]_{B_x B_x}$ and $[F_{\textrm{total (GHZ$_Z$)}}^{-1}]_{B_z B_z}$ have exactly the same form, but with different $f_1$ and $f_2$ functions.  These expressions show shot-noise scaling, unless $f_1 = 0$, which only happens in limiting cases (like $B_x = 0$).  For some values of $(B_x,B_z)$, $f_1$ will have small values.  Therefore, when we do numerical fittings of these expressions using small $N$ values, the $\beta$ exponent will be close to $2$ and we might incorrectly conclude that there is Heisenberg scaling.  However, as the $N$ range grows, the resulting $\beta$ approaches $1$.

\section{Ancilla-assisted protocol with dual GHZ-probe}

If we only use a probe in the $X$ basis $|\textrm{GHZ}_X \rangle \propto |+\rangle^{\otimes (N+1)} + |- \rangle ^{\otimes(N+1)}$, after inverting the CFIM, we get very similar expressions for the diagonal matrix elements of the inverse matrix:

\begin{equation}
    [F_{\textrm{total (GHZ$_X$)}}^{-1}]_{B_x B_x} = \frac{1}{4Nt^2} \left( f_1^{(xx, x)} + \frac{f_2^{(xx, x)}}{f_3^{(x)} + N} \right), \quad 
    [F_{\textrm{total (GHZ$_X$)}}^{-1}]_{B_z B_z} = \frac{1}{4Nt^2} \left( f_1^{(zz,x)} + \frac{f_2^{(zz,x)}}{f_3^{(x)} + N} \right) \,
\end{equation}

\begin{equation}
    \textrm{where} \quad f_1^{(xx,x)} = \frac{n_x^2n_z^2[2Bt -\sin(2Bt)]^2}{4\sin^2(Bt)[1-n_z^2\sin^2(Bt)]} \, \, , \, \, f_2^{(xx,x)} = \frac{[n_x^2+n_z^2Bt\cot(Bt)]^2}{1-n_z^2\sin^2(Bt)}
\end{equation}

\begin{equation}
    f_1^{(zz,x)} = \frac{[2n_x^2Bt + n_z^2\sin(2Bt)]^2}{4\sin^2(Bt)[1-n_z^2\sin^2(Bt)]} \, \, , \, \, f_2^{(zz,x)} = \frac{n_x^2n_z^2[1-Bt\cot(Bt)]^2}{1-n_z^2\sin^2(Bt)}, \quad f_3^{(x)} = \frac{n_z^2\sin^2(Bt)}{1-n_z^2\sin^2(Bt)}, 
\end{equation}

Now when we use two probes in unison, one in the $Z$ basis and one in the $X$ basis, we get the following analytical expressions for the inverted CFI matrix:

\begin{equation}
    [F_{\textrm{total (2 GHZ's)}}^{-1}]_{B_x B_x} = \frac{1}{4Nt^2} \left( \frac{n_x^2}{N+1} + \frac{2B_z^2 t^2}{A + CN} \right), \quad [F_{\textrm{total (2 GHZ's)}}^{-1}]_{B_z B_z} = \frac{1}{4Nt^2} \left( \frac{n_z^2}{N+1} + \frac{2B_x^2 t^2}{A + CN} \right), 
\end{equation}

 \noindent where $ A = \sin^2(Bt) [3-\cos(2Bt)], C = \sin^2(Bt) [1+\cos(2Bt)]$. These two expressions reveal a \textit{bona fide} Heisenberg scaling.  The only exception is when $C=0$, which occurs when $\cos(2Bt) = -1$.  This condition happens when $Bt=m\pi/2$ for every integer $m$.  The even $m$ values are the trivial cases (complete rotation of the probe and final state identical to initial state, so no information acquired).  The odd $m$ values are more interesting.  For the $|\textrm{GHZ} \rangle$ probe, when $m$ is odd, $B_{\textrm{eff}} = \arctan (n_z \tan(Bt))$ becomes $\pm \pi/2$.  This implies that the relative phase acquired by the probe after error correction becomes $\phi = \pm \pi (N-k)$ and $e^{i\phi} = \pm 1$.  The final state after error correction does not carry information about $B_x$ and $B_z$, apart from the fact that $Bt=m \pi/2$.  The Heisenberg scaling is lost but there is still shot-noise scaling if the information from the relative phase of the final corrected state is combined with the information from the stabilizer measurements (How many errors, $k$, were detected).The off-diagonal matrix elements have the following form:

\begin{equation}
    [F^{-1}]_{B_x B_z} = \frac{1}{4Nt^2} \left( \frac{f_1 + f_2 N}{f_3 + f_4N + f_5N^2} \right)
\end{equation}

\noindent Again, this expression reveals a true Heisenberg scaling, except when $f_5 = 0$.


\section{Generalization to three-dimensional magnetic field sensing}

The generalization of the previous results to a uniform, time-independent magnetic field with an extra $Y$ component turns out to be relatively straightforward.  Now, the Hamiltonian of our system is:

\begin{equation}
    H = B_x \sum_{j=0}^{N} \sigma_x^{(j)} + B_y \sum_{j=0}^{N} \sigma_y^{(j)} + B_z \sum_{j=0}^{N} \sigma_z^{(j)}, 
\end{equation}
where $\sigma_x^{(j)}$, $\sigma_y^{(j)}$, and $\sigma_z^{(j)}$ are the Pauli X, Y, and Z matrices acting on qubit $j$, respectively.  Setting $\hbar = 1$ and $t=1$, the time evolution operator is given by:

\begin{align}
    \hat{\mathfrak{U}} &= \exp(-i Ht) = [\exp(-it (B_x \sigma_x + B_y \sigma_y + B_z \sigma_z)) ] ^{\otimes N} \otimes I = [\cos(Bt) - i\sin(Bt) (n_x \sigma_x + n_y \sigma_y + n_z \sigma_z) ] ^{\otimes N} \otimes I
\end{align}

where $B {=} \sqrt{B_x^2 + B_y^2 + B_z^2}, n_x {=} B_x/B,n_y {=} B_y/B,n_z {=} B_z / B$. Like before, for the initial state $|\textrm{GHZ} \rangle_Z {=}  \left( |0\rangle^{\otimes (N+1)} {+} |1\rangle^{\otimes (N+1)} \right) / \sqrt{2}$, let the magnetic field act on every qubit except for the first one, and then measure the stabilizers of the $(N+1)$-qubit bit-flip code, the effective time evolution will be rotation of the original GHZ about the $Z$ axis.  The final state can be expressed in the following form below. Table \ref{tab:3D} summarizes all  the results.
\begin{equation}
    |\psi \rangle = \frac{1}{\sqrt{2}} \left(|0 \rangle ^{\otimes (N+1)} + e^{i \phi} |1 \rangle ^{\otimes (N+1)} \right) 
\end{equation}

\begin{table}[ht]
\centering
\renewcommand{\arraystretch}{1.5} 
\begin{tabular}{c|c|c}
\hline
Number of $X$ errors detected & Probability of occurrence & Relative phase ($\phi$) \\ \hline
$0$ & $(\cos^2(Bt) + n_z^2 \sin^2(Bt))^{N}$ & $2 B_{\textrm{eff}}^{(z)} N$ \\
$1$ & $\binom{N}{1} \left[ \cos^2(Bt) + n_z^2 \sin^2(Bt) \right]^{N-1} \left[ (n_x^2 + n_y^2)  \sin^2(Bt) \right]$ & $2 \left[ B_{\textrm{eff}}^{(z)} (N-1) + B_{\textrm{eff}}^{(x,y)} \right]$ \\ 
$\vdots$ & $\vdots$  & $\vdots$ \\
$k$ & $\binom{N}{k} \left[ \cos^2(Bt) + n_z^2 \sin^2(Bt) \right]^{N-k} \left[ (n_x^2 + n_y^2)  \sin^2(Bt) \right]^k$ & $2 \left[ B_{\textrm{eff}}^{(z)} (N-k) + B_{\textrm{eff}}^{(x,y)}k \right]$ \\
$\vdots$ & $\vdots$  & $\vdots$ \\
$N$ & $\left[ (n_x^2 + n_y^2)  \sin^2(Bt) \right]^{N}$ & $2 B_{\textrm{eff}}^{(x,y)}N$ \\ \hline
\end{tabular}
\caption{Probabilities of detecting $k$ ``errors'' and resulting relative phases after error correction for an ($N+1$)-qubit $|\textrm{GHZ}\rangle_Z$ probe and a magnetic field acting on all qubits except the first one.  The magnetic field $\vec{B} = (B_x,B_y,B_z)$ has $3$ components, $n_x=B_x/B$, $n_y=B_y/B$, $n_z=B_z/B$, and $B=|\vec{B}|=\sqrt{B_x^2+B_y^2+B_z^2}$. }
\label{tab:3D}
\end{table}

Now, we have two effective magnetic fields after error correction:
\begin{equation}
    B_{\textrm{eff}}^{(z)} = \arctan(n_z \tan(Bt)), \quad B_{\textrm{eff}}^{(x,y)} = \arctan(n_y / n_x).
\end{equation}

To obtain Heisenberg scaling, now we have to combine the information from $3$ probes:

\begin{align}
    |\textrm{GHZ}\rangle_Z &= \left( |0\rangle^{\otimes(N+1)} + |1\rangle^{\otimes(N+1)} \right) /\sqrt{2} \\
    |\textrm{GHZ}\rangle_X &= \left( |+\rangle^{\otimes(N+1)} + |-\rangle^{\otimes(N+1)} \right) /\sqrt{2} \quad, \quad |\pm\rangle = (|0\rangle \pm |1 \rangle) / \sqrt{2} \\
    |\textrm{GHZ}\rangle_Y &= \left( |+i\rangle^{\otimes(N+1)} + |-i\rangle^{\otimes(N+1)} \right) /\sqrt{2} \quad, \quad |\pm i \rangle = (|0\rangle \pm i|1 \rangle) / \sqrt{2}
\end{align}


\section{Bayesian analysis}

\begin{figure}
    \centering
    \includegraphics[width=0.7\linewidth]{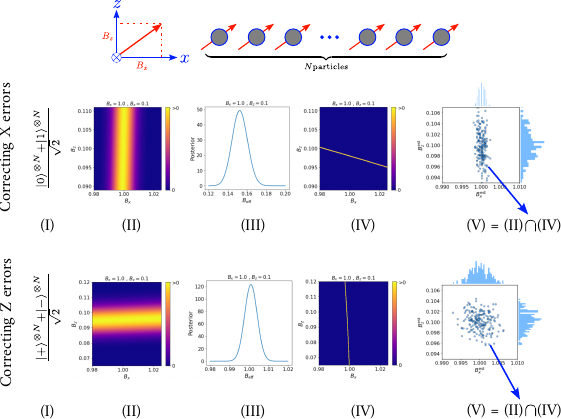}
    \caption{The top panel shows the quantum probe of Eq.~\eqref{eq_H_probe} used in our quantum error sensing scheme. The Bayesian procedure is as follows: Step (I), we select an appropriate probe to correct either $X$ or $Z$ errors. Step (II), we compute the posterior using the first likelihood, given in Eq.~\eqref{eq_first_likelihood}. Step (III), we compute the posterior using the second likelihood, given in Eq.~\eqref{eq_second_likelihood}. Step (IV), we use the theoretical expression for $B_{\textrm{eff}}$ together with the estimated value from the posterior in Step (III) to determine the set of $(B_x, B_z)$ values consistent with $B_{\mathrm{eff}}$. Step (V), we overlap the posterior from Step (II) with the set of possible values from Step (IV) to obtain the final refined estimate.}
    \label{fig_sm_QEC_bayesian}
\end{figure}

In this section, we clarify the Bayesian methods used in the ``Bayesian sensing protocol" section of the main text. As discussed throughout our work, the estimation procedure constitutes the final step in any quantum sensing protocol. We demonstrated that, using Bayesian analysis, our quantum error correction scheme can always achieve the Heisenberg limit of precision. In essence, the Bayesian methodology relies on Bayes' theorem:
\begin{equation}
\underbrace{P(\text{parameter}|\text{data})}_{\text{Posterior}}
=
\frac{
\overbrace{P(\text{data}|\text{parameter})}^{\text{Likelihood}} \hspace{5pt} \overbrace{P(\text{parameter})}^{\text{Prior}}
}{
\underbrace{P(\text{data})}_{\text{Normalization}}
}.
\end{equation}
In the above, the posterior function is the probability distribution over the parameter after observing the data. The prior encodes our previous knowledge or assumptions about the unknown parameter(s). The probability of the observed data serves purely as a normalization factor, ensuring that the posterior integrates (or sums) to one and thus remains a valid probability distribution.The likelihood function assigns a probability distribution from the observed data for a given value of the parameter and is the central quantity we evaluate in practice, as determined by our quantum statistical model. Importantly, the prior can be updated as new data are collected, which is why Bayes' rule acts as an updating rule, namely: the more data we acquire, the more sharply the posterior reflects the true value of the parameter. Recall that we employ two distinct GHZ probes, each targeting the correction of either $X$ or $Z$ errors (see main text). Consider, for example, the probe designed to correct $X$ errors. For this probe, we compute two posteriors: The first posterior is obtained from the information gained by measuring stabilizers. The second posterior is obtained from the final corrected state. Specifically, the first posterior is computed from the following likelihood function:
\begin{equation}
\text{First Likelihood} \to P(\text{Counting Observed Errors} | B_x,B_z) = \prod_k p_k^{\mathcal{C}_k}(B_x,B_z),\label{eq_first_likelihood}
\end{equation}
where $p_k(B_x,B_z)$ is the probability of correcting $k$ errors for a specific probe. Using $p_k(B_x,B_z)$, we simulate the outcomes of $M$ measurements, each indicating the number of errors detected. From this simulated data, we determine $\mathcal{C}_k$ as the total number of times exactly $k$ errors occur. The second likelihood is obtained from measurements on the fully corrected state. Specifically,
\begin{equation}
\text{Second Likelihood} \to P(\pm | B_{\mathrm{eff}}) = q_{k,+}^{\mathcal{C}_{k,+}} q_{k,-}^{\mathcal{C}_{k,-}},\label{eq_second_likelihood}
\end{equation}
where $q_{k,\pm}$ is the probability of obtaining a $+$ or $-$ outcome, respectively. Note that the label $k$ refers to the case of $k$ errors. However, this information is always accessible in a experiment and thus known once the state has been fully corrected. As above, $\mathcal{C}_{k,\pm}$ denotes the number of occurrences of each outcome over $M$ simulated trials. With both posteriors in hand, the final step is a discarding step, see Fig.~\ref{fig_sm_QEC_bayesian} for the step-by-step Bayesian procedure. In essence, the first posterior defines the region of parameter space where all plausible values of $B_x$ and $B_z$ lie. The second posterior provides additional information about the likely values of $B_{\mathrm{eff}}$. Armed with the estimated value of $B_{\mathrm{eff}}$, we determine the set of $(B_x, B_z)$ values consistent with $B_{\mathrm{eff}}$. By combining this information point-by-point and retaining only overlapping regions, we obtain the normalized posterior shown in Step (V) of Fig.~\ref{fig_sm_QEC_bayesian}. For a given probe, this entire procedure produces a single point in the scatter plot of Step (V) in Fig.~\ref{fig_sm_QEC_bayesian}. Repeating the experiment multiple times yields many such points, from which the covariance matrix can be directly evaluated. 

\end{document}